%% file: eprint.tex
\documentclass[10pt, paper=a4, UKenglish]{article}
\usepackage{graphicx}
\usepackage{float}
\def\Title#1{\begin{center} {\Large #1 } \end{center}}
\def\Author#1{\begin{center}{ \sc #1} \end{center}}
\def\Address#1{\begin{center}{ \it #1} \end{center}}

\newcommand\pubblock{\rightline{\begin{tabular}{l} Proceedings of the CTD 2023\\ \pubnumber\\
         \pubdate  \end{tabular}}}

\newenvironment{Abstract}{\begin{quotation} \begin{center} 
             \large ABSTRACT \end{center}\bigskip 
      \begin{center}\begin{large}}{\end{large}\end{center} \end{quotation}}

\newenvironment{Presented}{\begin{quotation} \begin{center} 
             PRESENTED AT\end{center}\bigskip 
      \begin{center}\begin{large}}{\end{large}\end{center} \end{quotation}}

\input econfmacros.tex

\usepackage{color}
\usepackage{lineno}
\usepackage{subfig}
\usepackage{hyperref}

\newcommand\pubnumber{PROC-CTD2023-43}

\newcommand\pubdate{\today}

\def\affiliation{
Lawrence Berkeley National Lab. \\
Cyclotron Rd., Berkeley, CA 94720, USA}
\def\affiliationn{
Department of Physics Engineering \\
Istanbul Technical University, Turkey}

\newcommand{\conference}{Connecting the Dots Workshop (CTD 2023)\\
October 10-13, 2023}

\usepackage{fancyhdr}
\definecolor{mygrey}{RGB}{105,105,105}
\begin{document}

\large
\begin{titlepage}
\pubblock

\vfill
\Title{An Application of HEP Track Seeding to Astrophysical Data}
\vfill

\Author{Mine G\"ok\c{c}en}
\Address{\affiliationn}
\Author{Maurice Garcia-Sciveres, Xiangyang Ju}
\Address{\affiliation}
\vfill

\begin{Abstract}
We apply methods of particle track reconstruction in High Energy Physics (HEP) to the search for distinct stellar populations in the Milky Way, using the Gaia EDR3 data set. This was motivated by analogies between the 3D space points in HEP detectors and the positions of stars (which are also points in a coordinate space) and the way collections of space points correspond to particle trajectories in the HEP, while collections of stars from distinct populations (such as stellar streams) can resemble tracks. Track reconstruction consists of multiple steps, the first one being seeding. In this note, we describe our implementation and results of the seeding step to the search for distinct stellar populations, 
and we indicate how the next steps will proceed. Our seeding method uses machine learning tools from the FAISS library, such as the k-nearest neighbors (kNN) search. 
\end{Abstract}

\vfill

\begin{Presented}
\conference
\end{Presented}
\vfill
\end{titlepage}
\def\thefootnote{\fnsymbol{footnote}}
\setcounter{footnote}{0}
%

\normalsize 

\section{Introduction} \label{intro}

The Milky Way (MW) was formed by the merging of many smaller objects and continues to accrete matter from satellite systems as it ages. Studying the progenitors of our galaxy is key to understanding its formation and evolution. Recent merger events are particularly useful in revealing insights about the the gravitational structure our galaxy, including its dark matter halo. As stars from satellite systems tidally get disrupted by our galaxy's gravitational potential, \textit{stellar streams} form around the galactic disk, inside its halo. Alongside with star clusters, they have been extensively studied by astronomers as relics of recent merger events. 

There have been numerous research efforts on identifying stellar streams and star clusters using statistical and computational techniques~\cite{Yuan_2018,streamfinder}. Although traditional methods of constraining stars by their color, age and kinematics are still in use~\cite{Chandra_2023}, simulations~\cite{Shipp_2023, Bonaca_2014} are widely utilized to determine star candidates for discoveries of these astrophysical structures. More novel methods of anomaly detection using machine learning (ML)~\cite{Pettee:2023zra} and deep learning~\cite{Shih:2023jfv} are also being actively investigated with the common purpose of automating searches for distinct stellar populations.   

Orbits of stellar streams are reminiscent of particle tracks in high energy physics (HEP) detectors. In this work, we explore applying (HEP) track reconstruction methods to astrophysical data, inspired by that resemblance, and more generally to identify stars from distinct populations in an automated way. 

In HEP experiments, the trajectory a particle follows inside the detector needs to be reconstructed—a process called particle tracking—in order to study the particle interactions and events that have occurred inside. In essence, tracking consists of three major steps: forming seeds, extrapolating the seeds to form track candidates, and fitting tracks using these seeds. Particle detectors have multiple coaxial layers that record the location a particle passes by. These locations that are marked when a particle hits the detector layer are called space-points (SP). Seeds are clustered sets of SPs that potentially belong to the same particle track, which are then combined and extended to arrive at final track candidates. 

We aim to establish an analogous method to track stars—tracking meaning identifying stars with a common origin in this context. Stars belonging to a certain stellar population could be characterized by their intrinsic properties, such as their ages and chemical compositions, as well as their kinematics. Here, to construct seeds of stars with shared origins, we will cluster them based on their kinematics in the sky. In a later step we would extend and combine the seeds to arrive at final distinct populations. 

This paper describes our ML-based seeding method as the first stage of tracking stars. In section~\ref{dataset}, we introduce the astrophysical data set used for our analysis. Section~\ref{methods} goes over the seeding method. In section~\ref{results}, we share our current seed results and give examples of known objects that collections of seeds clearly correspond to. Finally, in section~\ref{discuss}, we conclude with a discussion on the performance of our seeding method and the steps for implementing the second tracking stage to automatically extend and combine seeds.   

\section{Gaia Dataset} \label{dataset}

\begin{figure}[H]
  \centering
  \subfloat[]{\includegraphics[width=0.48\linewidth]{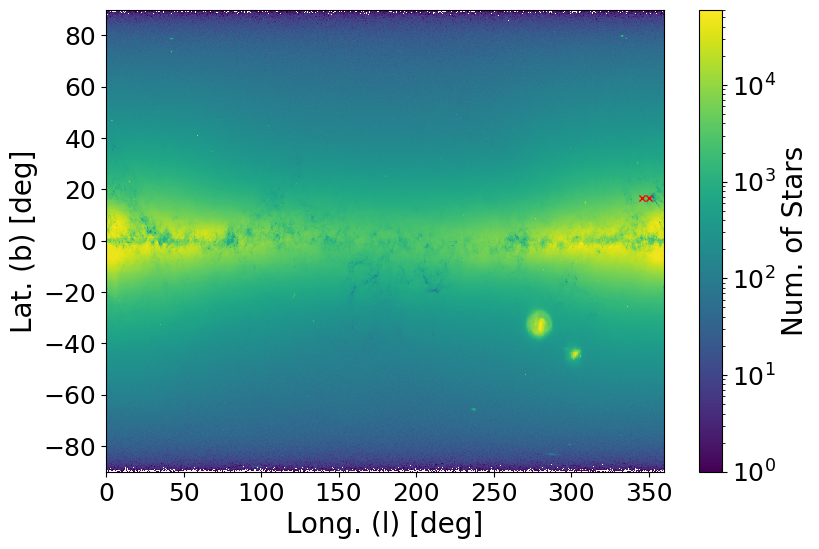}}
  \quad
  \subfloat[]{\includegraphics[width=0.48\linewidth]{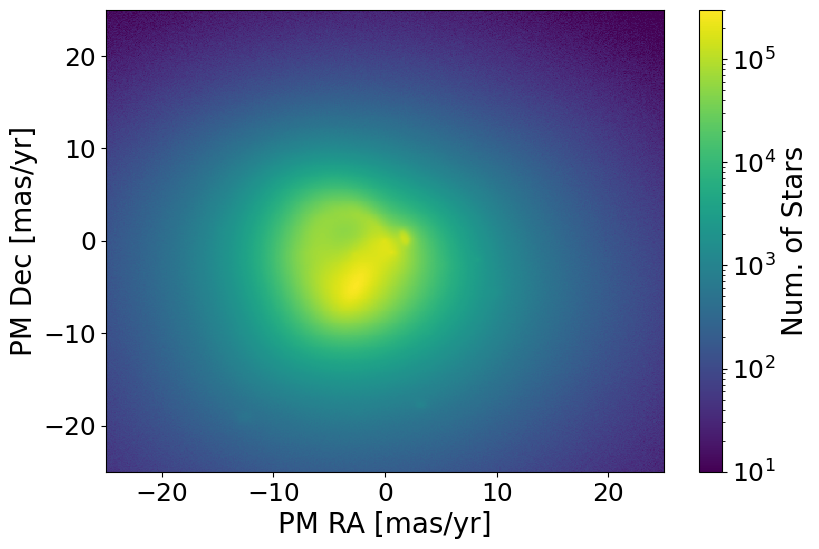}}
  \caption{All Gaia EDR3 stars with parallax (total of 1,467,744,818 stars) binned in (a) position- and (b) proper motion-space, with respective bin sizes of 0.25 deg and 0.1 mas/yr in each axis. The two small red crosses at (l, b) $\sim$ (350, 20) degs in (a) mark the patches of sky that will be shown in fig.~\ref{fig: clustering} and discussed in section \ref{methods}.}
  \label{fig: all stars}
\end{figure}

We use the EDR3 catalog~\cite{edr3} of the \textit{Gaia} mission~\cite{gaia}. Gaia EDR3 has observations of roughly two billion stars and their properties. In particular, we use the following five positional and kinematic parameters: galactic longitude and latitude,  proper motion (PM) in right ascension and declination, and parallax. It is important to note that parallax is a parameter with high uncertainties \cite{parallax} and could yield unreliable results if not treated carefully. Therefore, we mainly use the first four parameters in our seeding method. The distribution of catalog stars in the 2D position and kinematics space is illustrated in fig.~\ref{fig: all stars}.

\section{Seeding Methods} \label{methods}

\subsection{Subsetting the Data}

The seeding will look for differences in PM as a signature of distinct populations. The basic assumption of our seeding method is that most stars have a PM given by the galaxy average, which we call background stars, while a fraction of stars that originated from recent mergers or accretion (signal stars) will move differently. Additionally and critically, we assume each signal population is localized in one part of the sky. Therefore, if the whole sky PM distribution were considered at once (fig.~\ref{fig: all stars}(b)), the correlation between PM and position would be hidden.  

On the other hand, in a small patch of sky (from here on referred to as a \textit{chunk}), signal stars have a chance to stand out in PM space. Thus, we carry out PM space clustering of stars chunk-by-chunk, rather than globally.

\begin{figure}[H]
    \centering
    \subfloat[]{\includegraphics[width=0.46\linewidth]{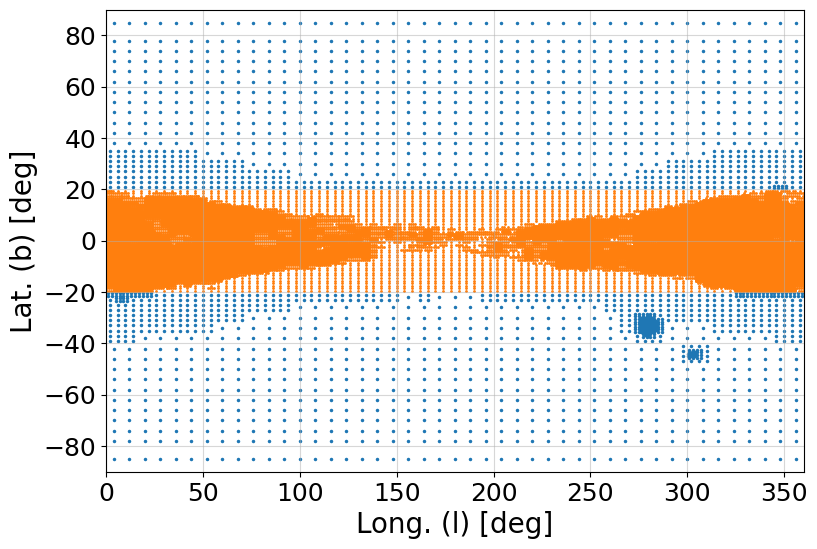}}
    \qquad
    \subfloat[]{\includegraphics[width=0.46\linewidth]{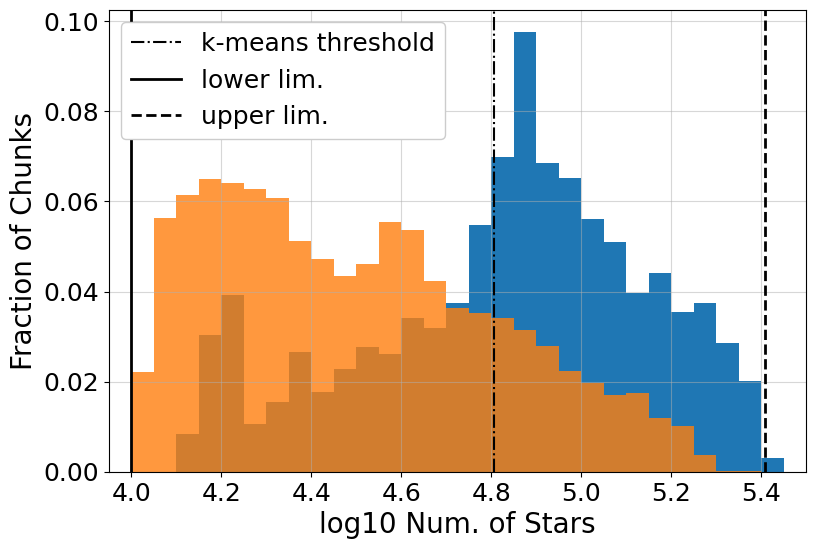}}
    \caption{Gaia stars subsetted in the position space. (a) shows position-space centers of the chunks and (b) is the binned number of stars in chunks. Orange represents the galactic disk region ($\sim$ 30,500 chunks) and blue represents the galactic halo region ($\sim$ 2,500 chunks).}
    \label{fig: chunks}
\end{figure}

Chunking, or subsetting stars in position-space is not as trivial as using a fixed grid. The number of stars in a chunk can have a considerable effect on the clustering results, depending on clustering parameters. We pick the minimum and maximum number of stars to be contained in a chunk based on a sample size threshold that the clustering algorithm we use is optimized for. The chunk sizes vary within these limits and the distribution of number of stars in chunks is shown in fig.~\ref{fig: chunks}(b) along with centers of the chunks in fig.~\ref{fig: chunks}(a). Note that we make no attempt to match the chunk size to the expected physical extent of some distinct population or another, because we are only searching for seeds at this stage, which can be thought of as hints of where distinct populations may reside in both PM and position. Eventually, when populations are fully reconstructed, there will be no memory of the seeds and chunks, just as there is no memory of the seeds in the HEP final reconstructed tracks.  

Because the density of stars in the galactic disk region ($\mid$b$\mid \leq$ 20 degrees) is much higher than that of the halo region ($\mid$b$\mid > $ 20 degrees), the chunk dimensions in these two regions differ significantly. Analysis of seed results also requires subtraction of extra-galactic backgrounds in the halo region; so disk and halo seeds must be analyzed separately. Due to space limitations, we focus the rest of this paper on galactic disk region. 

\subsection{Clustering Stars}

\begin{figure}[!htb]
  \centering
  \subfloat[]{\includegraphics[width=0.46\linewidth]{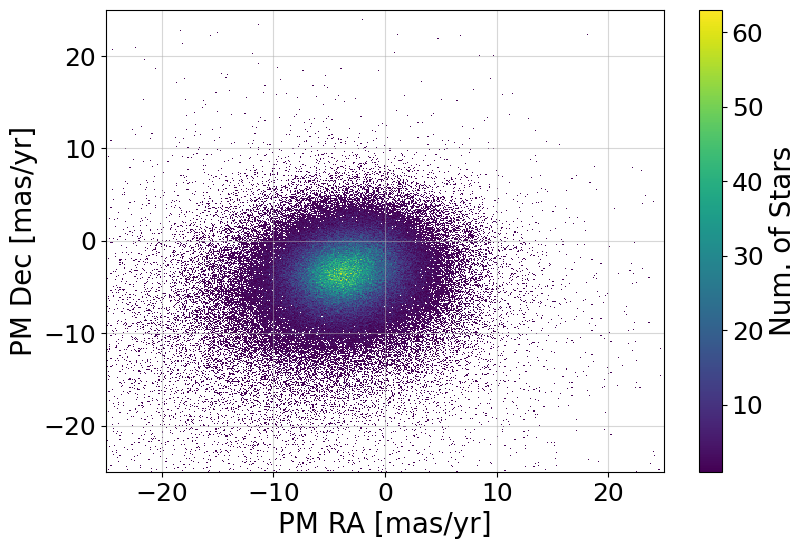}}
  \qquad
  \subfloat[]{\includegraphics[width=0.46\linewidth]{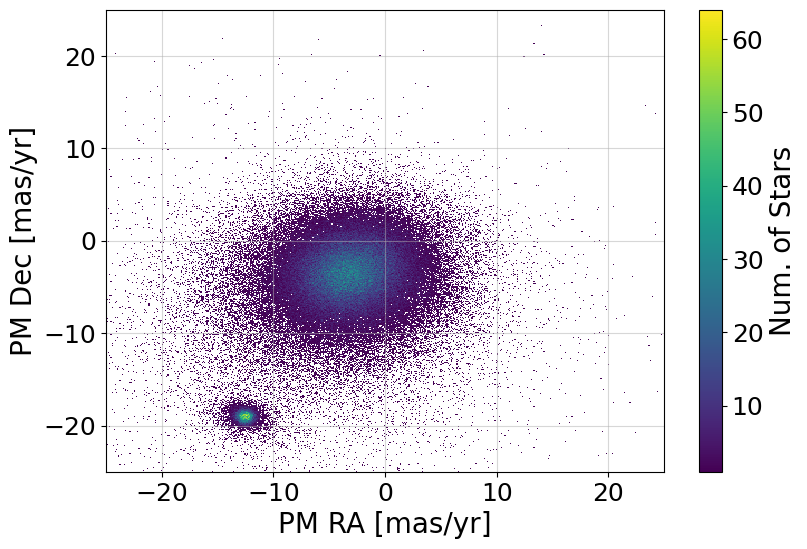}}
  \quad
  \subfloat[]{\includegraphics[width=0.46\linewidth]{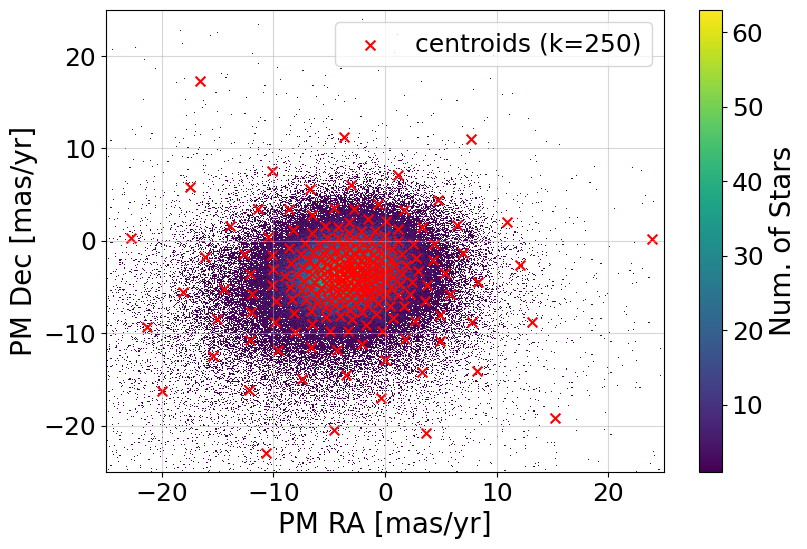}}
  \qquad
  \subfloat[]{\includegraphics[width=0.46\linewidth]{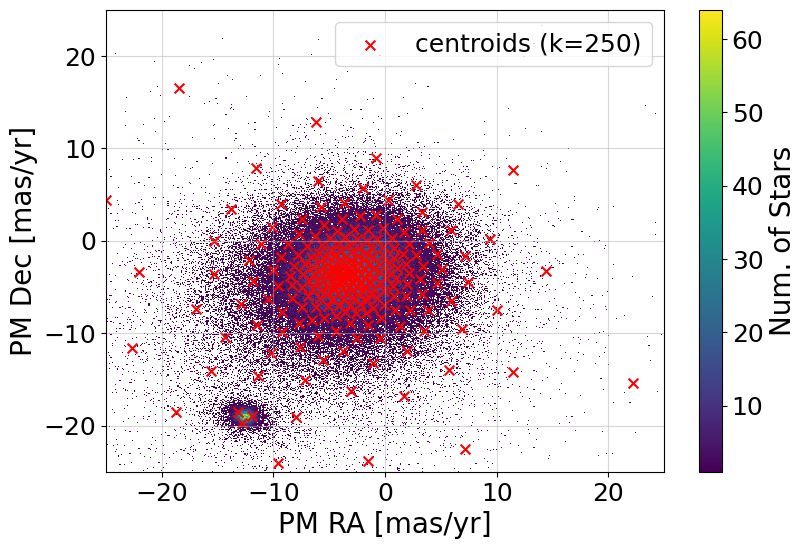}}
  \caption{The stars from  two exemplary neighboring chunks
  marked by red crosses in fig.~\ref{fig: all stars}(a)
  are shown in PM space in (a) and (b). Each chunk spans (dl, db) $\sim$ (4, 1) degrees in position space. Chunk (a) contains only background stars, whereas (b) has a secondary peak of signal stars. Panels (c) and (d) show the same stars with the 250 centroids found by k-means clustering with k=250 overplotted.}
  \label{fig: clustering}
\end{figure}

Clustering is done in each chunk individually based on the two dimensional PM of stars. 
Fig.~\ref{fig: clustering} shows the PM distributions in two selected neighboring chunks, which have been singled out in fig.~\ref{fig: all stars}(a). A very clear signal can be seen in 
fig.~\ref{fig: clustering}(b). Note that there is no signal in the chunk just next door (a), which illustrates the point that signals can have high PM-position correlation.    
The goal of clustering is to find such signal peaks in an automated way; not only the obvious ones ones like in fig.~\ref{fig: clustering}(b), but also ones that may be hard to spot visually. 

We cluster stars using the open source \textbf{FAISS} library~\cite{faiss}, which can conduct the clustering on GPU. More specifically, we use the FAISS k-means algorithm to obtain suggested centroids for a given number of clusters, $k$, and the k-NN algorithm to uniquely assign each star in the chunk to one of the $k$ centroids. 
Fig.~\ref{fig: clustering}(c) and (d) show the results of clustering with $k=250$ on (a) and (b), respectively. 

If all stars in a chunk have a background distribution, which has a broad shape as in  fig.~\ref{fig: clustering}(a), 
then the number of stars in every centroid will be approximately $N/k$ plus or minus statistical fluctuations, where $N$ is the number of stars in the chunk. 
If a signal is present, which is more peaked than the background, the centroids containing the signal will have a number of stars determined by the signal density, and not by $N/k$. This makes it clear that $1/k$ plays the role of a signal detection threshold.   

\subsection{\label{sec:seeds} Testing for Seeds}

\begin{figure}[H]
    \centering
    \subfloat[]{\includegraphics[width=0.475\linewidth]{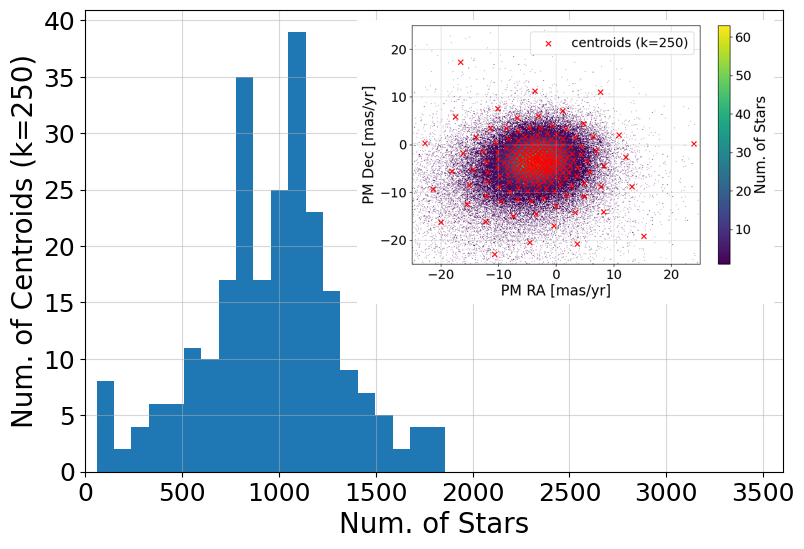}}
    \qquad
    \subfloat[]{\includegraphics[width=0.475\linewidth]{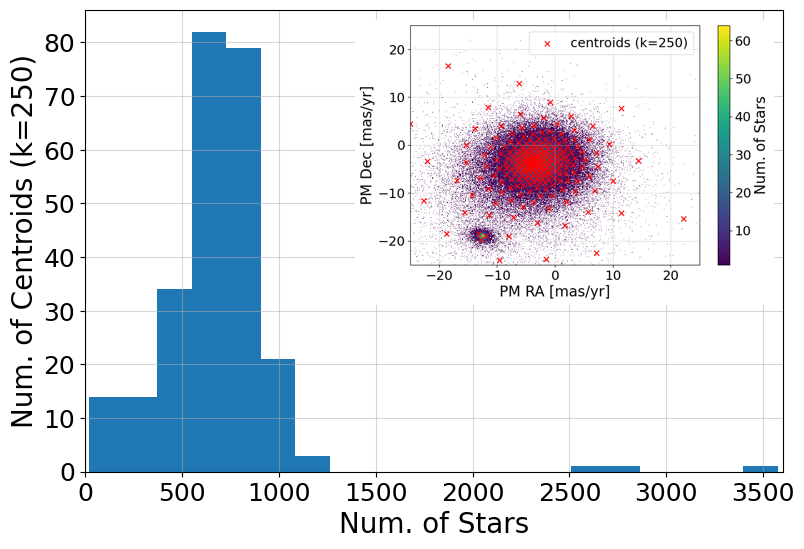}}
    \caption{Distribution of number of stars per centroid in k-means clusters for k=250, with the inserts showing the cluster centroids in PM space (crosses). Plot (a) shows the chunk of fig.~\ref{fig: clustering}(a), which failed the seed test (no seed detected in this chunk), while (b) is that of fig.~\ref{fig: clustering}(b), which passed seed test. The seed stars can be clearly seen in the insert and as well separated bins on the right side of the histogram.}
    \label{fig: test}
\end{figure}

After we perform the clustering over each chunk, we automatically test if a chunk contains a seed by checking for outliers in the binned distribution of number of stars per centroid, as shown in fig.~\ref{fig: test}. 
We look for high-side gaps between the bins, and cut on the counts within the outlier bins. We empirically settled on using 20 bins, having checked that varying between 15 and 25 bins made little difference. We investigated more formal statistical methods and found that this simple test was the most robust against differences in the shape of the main distribution. As in HEP track reconstruction, seeding does not have to be perfect, as seeds are intended to be redundant hints to be followed up with extension and merging later. In fig.~\ref{fig: test}, example (a) fails the seed test (no signal found) while example (b) passes. 

The simple gap selection method works well for small signals (where most of the stars in the chunk are not signal), but can fail when the number of signal stars in a chunk approaches or exceeds the number of background stars, in which case the histogram in number of stars per centroid can have two peaks with no gap in between. We have not added a test to detect those extreme seeds, because such a large signal will spread over many chunks and will already have many redundant seeds. 
As in HEP track reconstruction, one needs a sufficient number of seeds to cover the objects to be reconstructed, and adding more redundant seeds will not change the outcome. 

Once a chunk passes the seed test, we save the stars within the top three k-means clusters with highest number of stars as seed stars. One reason for saving multiple cluster stars is the possibility of having more than one signal peak in a chunk. Another is that there are often more than one centroid per signal peak. 

\begin{figure}[H] 
  \centering
  \subfloat[]{\includegraphics[width=0.31\linewidth]{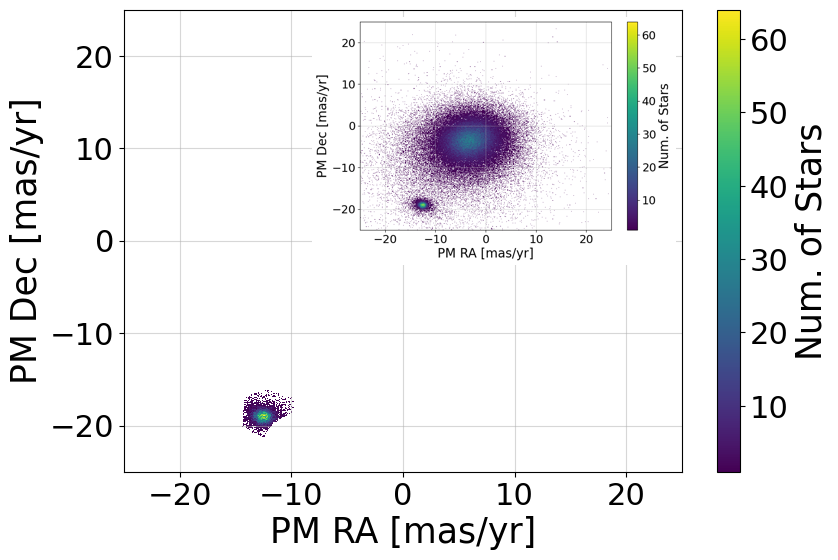}}
  \quad
  \subfloat[]{\includegraphics[width=0.31\linewidth]{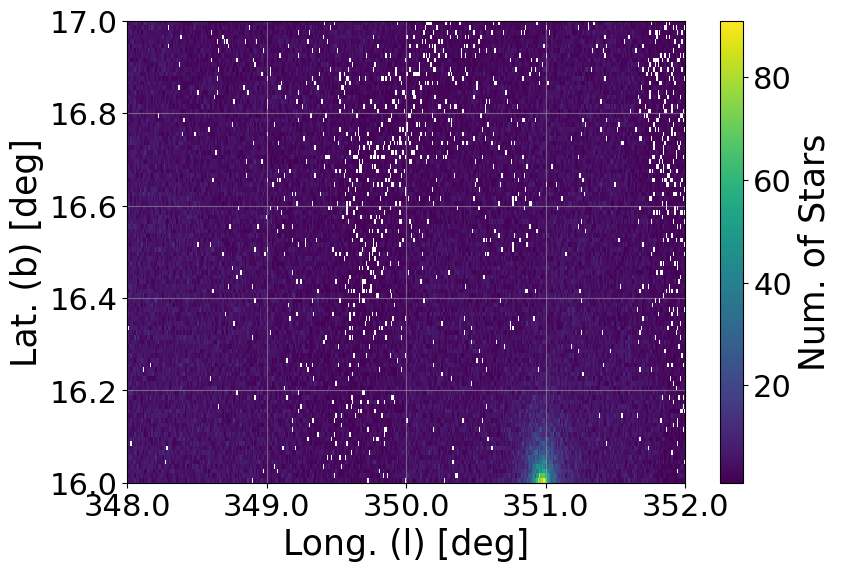}}
  \quad
  \subfloat[]{\includegraphics[width=0.31\linewidth]{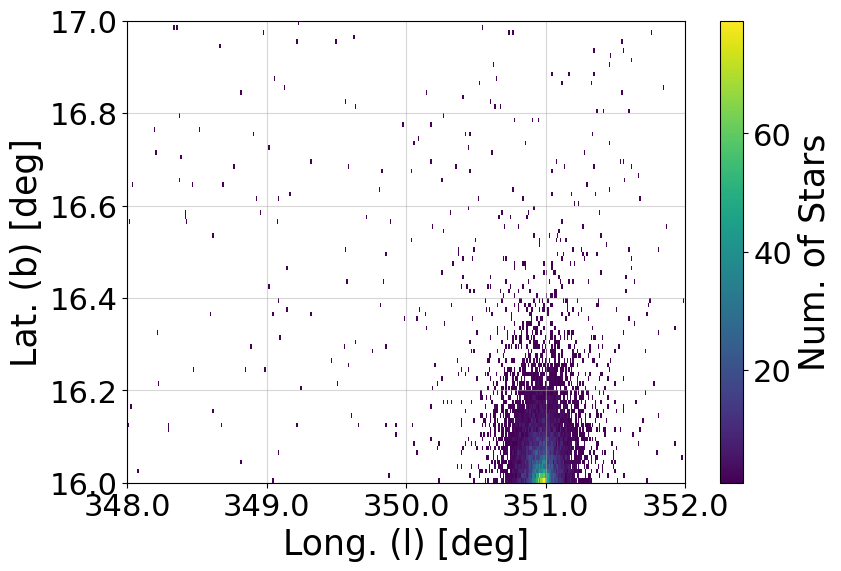}}
  \caption{Example of a seeded chunk. All chunk stars were plotted in fig.~\ref{fig: clustering} (reproduced here in the insert of (a)). The seed stars are shown in PM space in (a) and position space in (c), while  (b) shows all chunk stars in position space.}
  \label{fig: seeded}
\end{figure}

\begin{figure}[H] 
  \centering
  \subfloat[]{\includegraphics[width=0.31\linewidth]{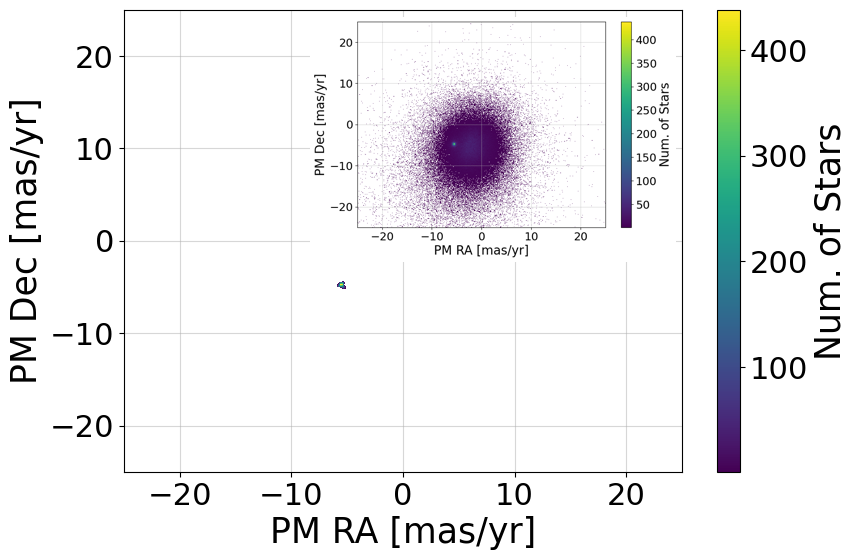}}
  \quad
  \subfloat[]{\includegraphics[width=0.32\linewidth]{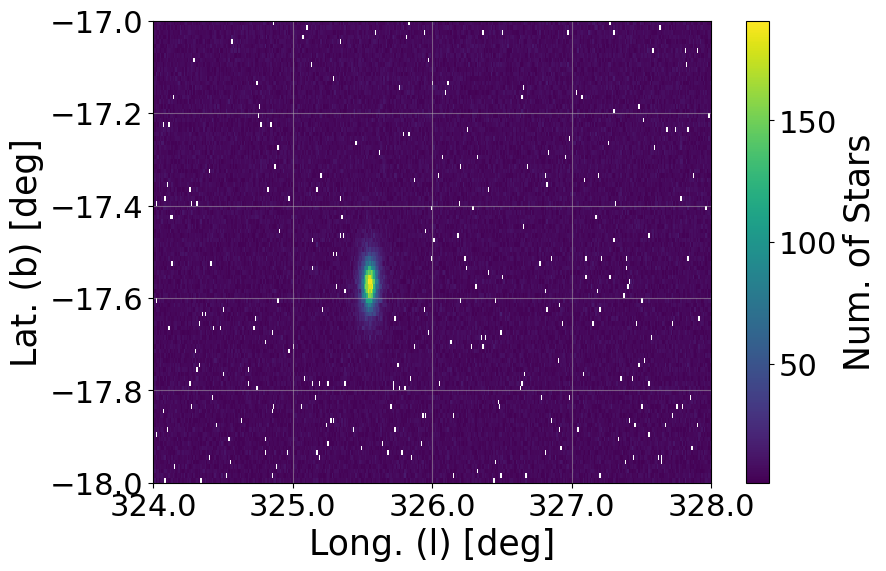}}
  \quad
  \subfloat[]{\includegraphics[width=0.31\linewidth]{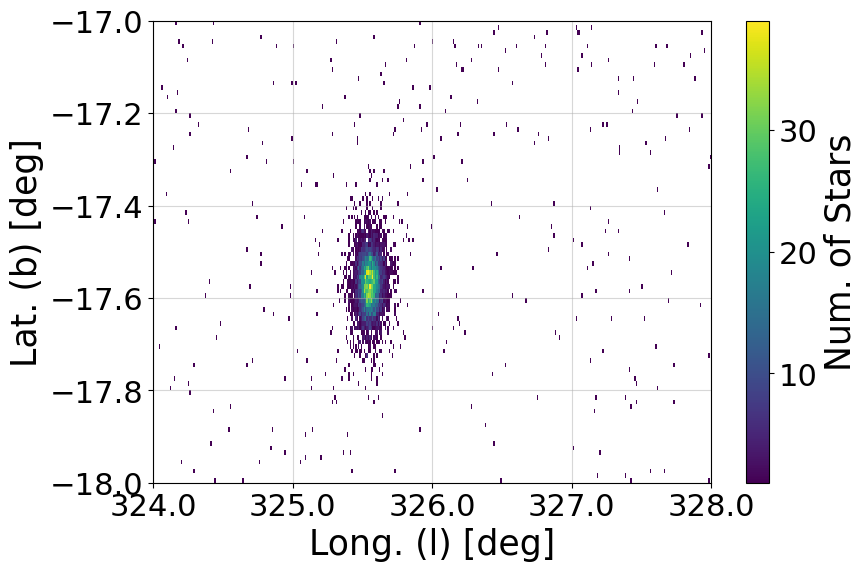}}
  \caption{Another example of a seeded chunk. The insert in (a) shows all chunk stars in PM space, 
  while (a) shows only the seed stars. (b) Shows all
  chunk stars in position space while (c) shows only the seed stars.}
  \label{fig: seeded 2}
\end{figure}

Fig.~\ref{fig: seeded} shows how seed stars identified by our method in PM space also cluster in the position space, as expected. A second example in which the signal peak overlaps the background distribution in PM space is illustrated in fig.~\ref{fig: seeded 2}. 

\section{Seed Results} \label{results}

\begin{figure}[H]
    \centering 
    \subfloat[]{\includegraphics[width=0.48\linewidth]{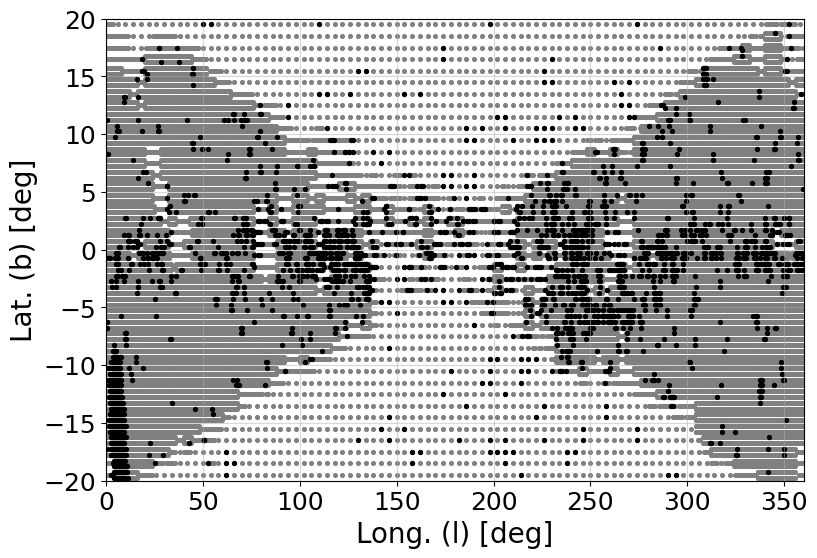}}
    \quad
    \subfloat[]{\includegraphics[width=0.48\linewidth]{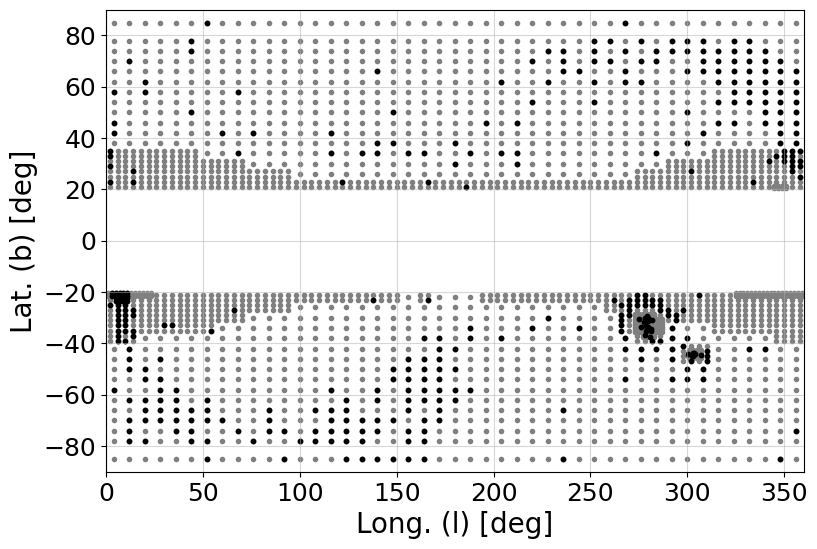}}
    \caption{Seed results in position space, where gray points are all chunk centers and black represents the seeded ones. (a) is that of galactic disk region and (b) is that of galactic halo region. In (a), all seed results are shown, including the seeds from chunks with offset centers, with the corresponding original chunk centers marked for visualization purposes.}
    \label{fig: seeds-pos}
\end{figure}

\begin{figure}[H]
    \centering
    \subfloat[]{\includegraphics[width=0.48\linewidth]{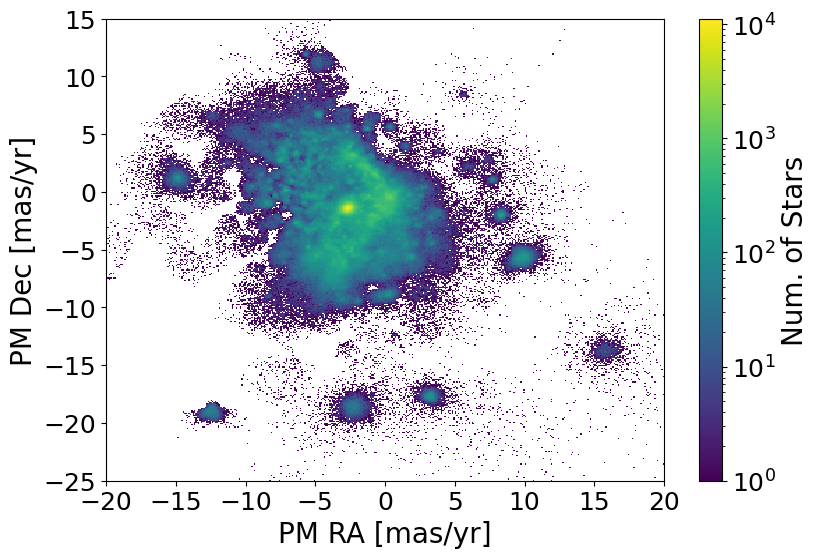}}
    \quad
    \subfloat[]{\includegraphics[width=0.48\linewidth]{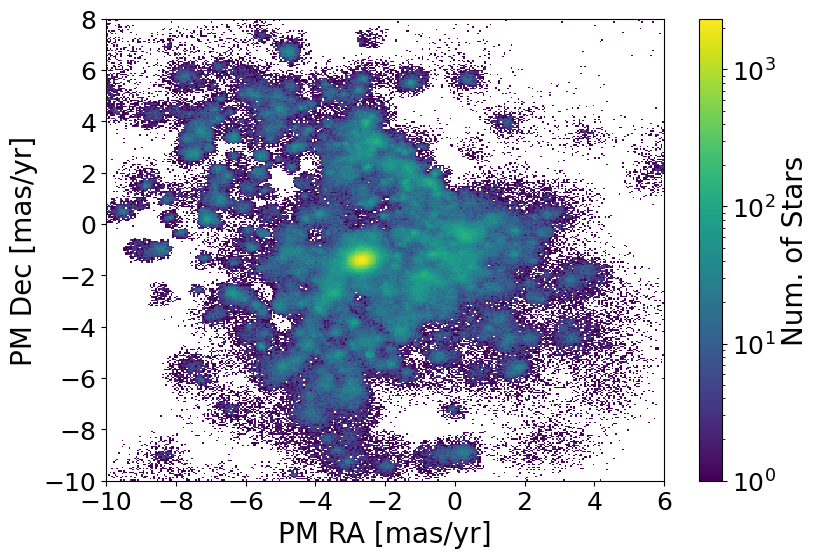}}
    \caption{Binned galactic disk seed stars in PM space: (a) shows all seed stars and (b) zoomed in central region showing only the seed stars from the most populated k-means clusters in seeded chunks (as opposed to the top three).}
    \label{fig: seeds-pm}
\end{figure}

This section shows the seeding results for the galactic disk region. Fig.~\ref{fig: seeds-pos} and fig.~\ref{fig: seeds-pm} show the results in position and PM space respectively, with statistics given in table~\ref{tab: results}. The halo region still requires the second chunking step described below and the removal of extra-galactic backgrounds that will cluster near zero proper motion (i.e. fixed in the sky), because they are very distant objects. 

\begin{table}[H]
  \begin{center}
    \begin{tabular}{c|c|c|c|c}
      \hline
      \hline
      & Total Chunks & Seeded Chunks & Percentage of Seeded Chunks & Seeded Stars \\
      \hline
      Galactic disk & 30,264 & 2,554 & 8.44 & 1,971,745 \\
      \hline
      Galactic halo & 2,375 & 618 & 26.0 & 354,557 \\
      \hline
      \hline
    \end{tabular}
    \caption{Overview of galactic disk (top row) and preliminary galactic halo (bottom row) seed results. Galactic halo results are only preliminary: they are missing the addition of seeds from chunks with offset centers. Seeds corresponding to extra-galactic backgrounds with zero proper motion (i.e. fixed in the sky because they are very distant) are removed.}
    \label{tab: results}
  \end{center}
\end{table}

For the galactic disk region, we performed the chunking twice. Stellar populations residing at the corners of chunks may not have enough signal stars in any chunk and thus, tend to lack seeds. Shifting chunk centers diagonally such that the corners of original chunks become the centers for new ones accounts for this vulnerability. For each set of chunks, we ran our seeding algorithm for $k=250$ and $k=600$, which amounts to four runs of seeding in total. At the end, seed stars from all runs are combined and get checked for redundancy.

Extending (meaning adding non-seed stars) and combining seed stars into candidate objects is the next step in the HEP-inspired reconstruction chain and the automation of this step is work in progress. Such object candidates 
can be compared to astrophysical object catalogs. 
We give two examples of known objects that match seed star collections, where the collection was done manually. 

\begin{figure}[H] 
  \centering
  \subfloat[]{\includegraphics[width=0.30\linewidth]{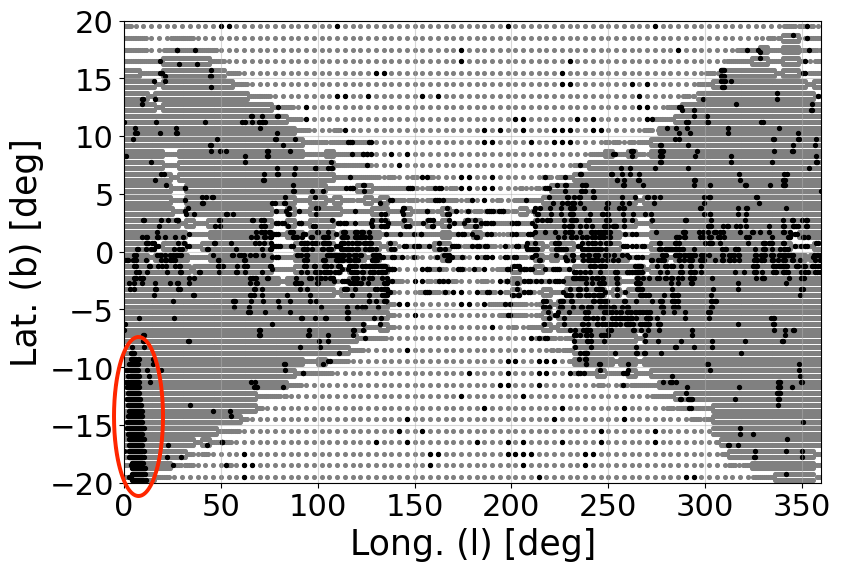}}
  \quad
  \subfloat[]{\includegraphics[width=0.31\linewidth]{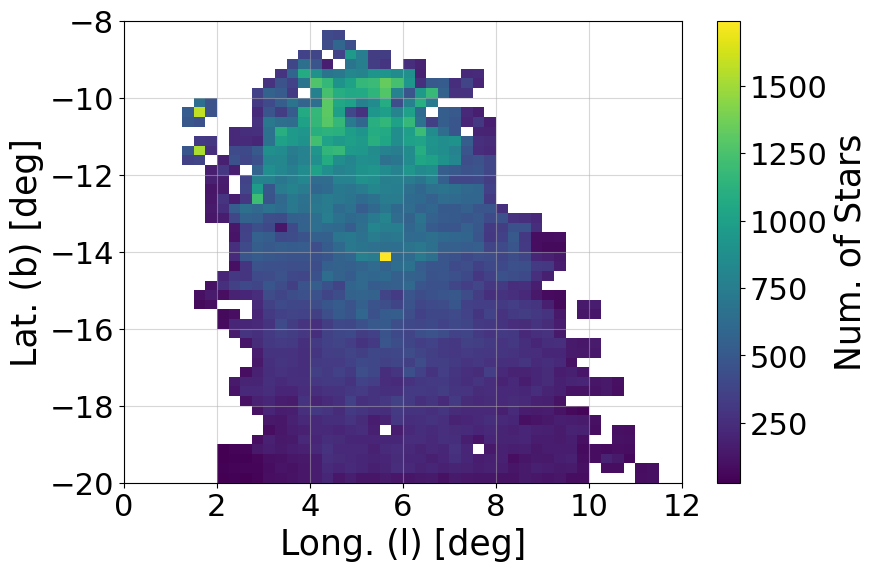}}
  \quad
  \subfloat[]{\includegraphics[width=0.31\linewidth]{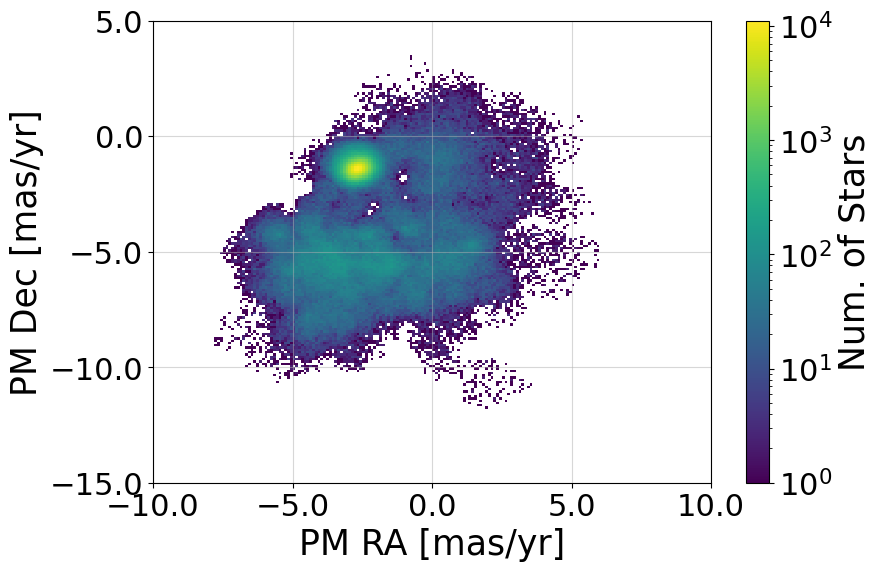}}
  \caption{Seed stars and seeded chunks matching the Sagittarius dwarf galaxy and stream. In (a), associated seeded chunk centers are circled with red. In (b) position space and in (c) PM space distributions are shown for the seed stars within the selection in (a).}
  \label{fig: sagittarius}
\end{figure}

The first example is the Sagittarius dwarf galaxy~\cite{sagit}, which stands out due to  the many neighboring seeded chunks circled in fig.~\ref{fig: sagittarius}(a). We compare the PM of seed stars in these chunks, plotted in fig.~\ref{fig: sagittarius}(c), along with their galactic coordinates and parallax, to the SIMBAD data base~\cite{simbad} objects and find they are well matched. 

\begin{figure}[H] 
  \centering
  \subfloat[]{\includegraphics[width=0.30\linewidth]{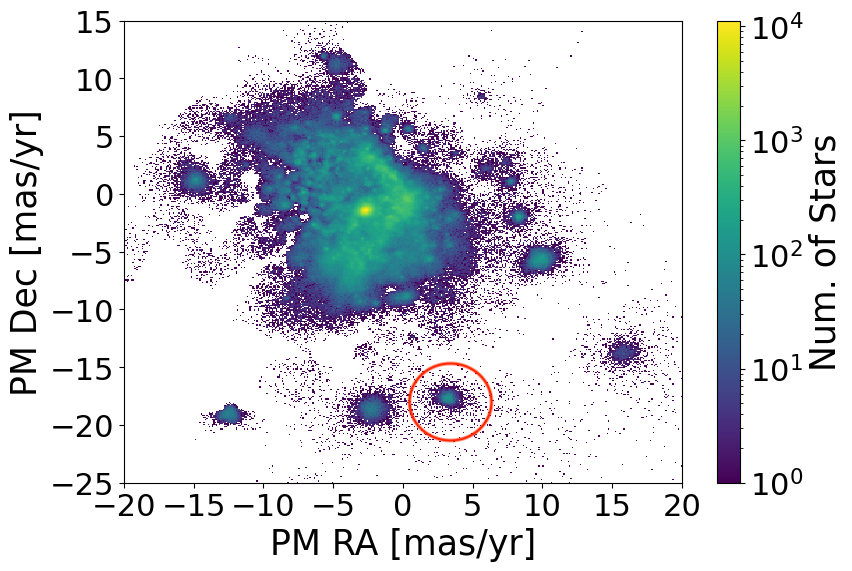}}
  \quad
  \subfloat[]{\includegraphics[width=0.31\linewidth]{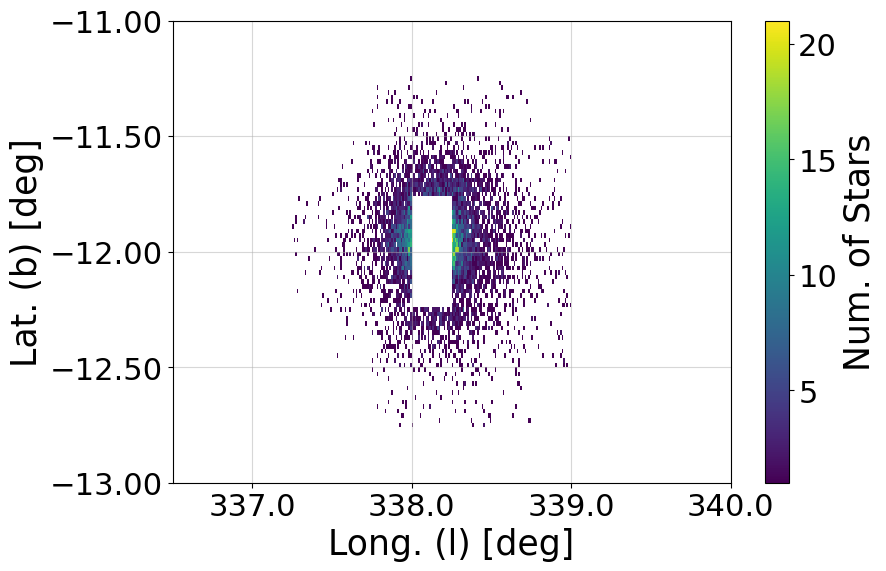}}
  \quad
  \subfloat[]{\includegraphics[width=0.30\linewidth]{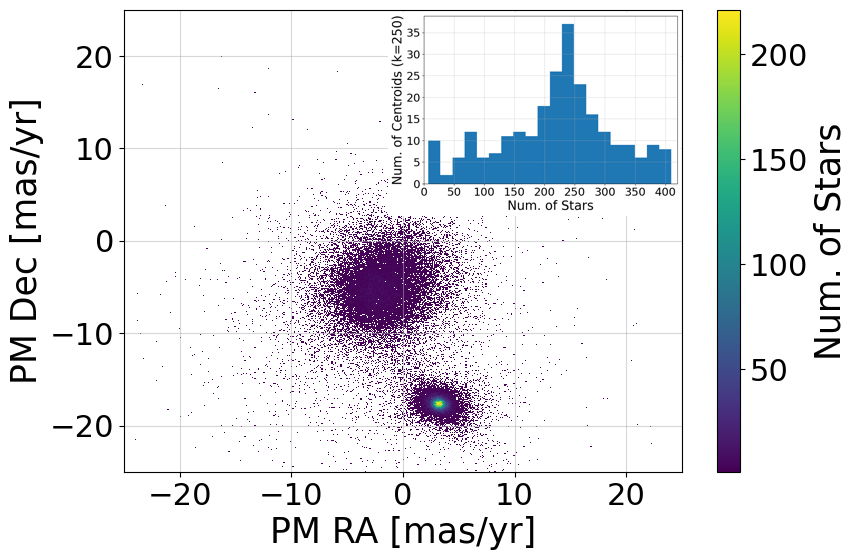}}
  \caption{Seed stars matching the known NGC 6397 globular cluster. In (a), associated seed stars are circled with red among all the galactic disk seed stars. In (b), seed stars are shown in position space. The central chunk in (b), whose PM space distribution is shown in (c), failed the seed test due to an extremely high signal content, as explained in the text. }
  \label{fig: ngc6397}
\end{figure}

The second example is the NGC 6397 globular cluster~\cite{ngc}, which matches the distinguishable PM-space distribution of its associated seed stars circled in fig.~\ref{fig: ngc6397}(a). The position-space center of these seed stars, plotted in fig.~\ref{fig: ngc6397}(c), along with their PM and parallax center, indicate that these are potentially NGC 6397 stars. The gap in the position space region of NGC 6397 seed stars is due to the extreme signal strength in that chunk, as explained in sec.~\ref{sec:seeds}.

\section{Conclusion and Future Work} \label{discuss}

We established an ML-based, automated seeding method to identify distinct stellar populations. Seeding is the first step of a HEP track reconstruction inspired object search. We tested our method on Gaia EDR3 stars and presented our seed results for the galactic disk region.

The next stage after seeding, analogous to HEP track reconstruction, is under development. For this purpose, we have begun exploring methods to combine and extend seeds as well as to refine them. Re-clustering seed stars in position and PM space stands out as a promising start. However, we face challenges due to the varying structures and sizes of stellar populations in these parameter spaces. 

This work suggests that it is possible to develop an automated reconstruction flow that, starting from survey data, can potentially identify in an instant all known stellar populations that have been manually cataloged over centuries of varied observations. If that is the case, previously unknown populations could also be identified. Significant further work is needed to reach this point.   

\section{Acknowledgements}
We thank Jessica Lu for helpful discussion. This work was supported in part by U.S. Department of Energy, Office of Science under contract DE-AC02-05CH11231.

This work has made use of data from the European Space Agency (ESA) mission
{\it Gaia} (\url{https://www.cosmos.esa.int/gaia}), processed by the {\it Gaia}
Data Processing and Analysis Consortium (DPAC,
\url{https://www.cosmos.esa.int/web/gaia/dpac/consortium}). Funding for the DPAC
has been provided by national institutions, in particular the institutions
participating in the {\it Gaia} Multilateral Agreement.


\end{document}

%% file: econfmacros.tex
\def\beq{\begin{equation}}
\def\eeq#1{\label{#1}\end{equation}}
\def\eeqn{\end{equation}}

\def\beqa{\begin{eqnarray}}
\def\eeqa#1{\label{#1}\end{eqnarray}}
\def\eeqan{\end{eqnarray}}

\let\bar=\overbar

\def\Dslash{\not{\hbox{\kern-4pt $D$}}}
\def\dslash{\not{\hbox{\kern-2pt $\del$}}}

\def\msb{{\bar{\ssstyle M \kern -1pt S}}}

